# Optical, luminescence, and scintillation properties of ZnO and ZnO:Ga ceramics

E. I. Gorokhova,[a)] G. V. Anan'eva, and V. A. Demidenko

*Scientific Research and Technological Institute of Optical Material Science, S. I. Vavilov State Optical Institute All-Russia Science Center, St. Petersburg*

P. A. Rodnyĭ and I. V. Khodyuk

*St. Petersburg State Polytechnical University, St. Petersburg*

E. D. Bourret-Courchesne

*Lawrence Berkeley National Laboratory, University of California, Berkeley, USA*

Uniaxial hot pressing has been used to obtain ceramics based on zinc oxide, and their optical, x-ray-structure, luminescence, and scintillation characteristics have been studied. It is shown that, by changing the concentration of the dopant (Ga) and the codopant (N), it is possible to change the intensities of the edge band (397.5 nm) and the intraband luminescence (510 nm) of the ZnO luminescence, as well as their ratio. Undoped ZnO ceramic has good transparency in the visible region and fairly high luminous yield: 9050 photons per MeV. Ceramic ZnO:Ga possesses intense edge luminescence with a falloff time of about 1 ns. © *2008 Optical Society of America.*

## INTRODUCTION

Powdered zinc oxide doped with gallium possesses a short de-excitation time of 0.7 ns and a high luminous yield of 15 000 photon/MeV.[1] As a result, ZnO:Ga has the highest quality (the ratio of the luminous yield to the falloff time) among known phosphors. Good scintillation characteristics are also shown by the phosphors ZnO:In (Refs. 1 and 2) and ZnO:Zn.[3,4] However, powdered and thin-film materials are used only for detecting neutrons and alpha particles,[4,5] whereas scintillators that possess a large volume (single crystals or optical ceramics) and high transparency in the spectral emission region of the material are needed for recording gamma and x-ray quanta. The production of bulk, single-crystal ZnO is a difficult, time-consuming, and expensive technological process. There are at present only isolated reports of the growth of ZnO-based single-crystal scintillators, in particular ZnO:In, of millimeter dimensions.[2]

A scintillation ceramic, which is a pressed polycrystalline material whose grain size can vary within wide limits, is a promising alternative to traditional scintillation single crystals. Crystals of cubic syngony are most acceptable for producing optical ceramics (ceramics that are transparent to their own radiation).[6] It is much harder to obtain a transparent ceramic in the visible region in the case of crystals with low symmetry (such as ZnO), since microstructure is formed from the disoriented grains during the hot pressing of an optical ceramic, and this causes significant light scattering even when the refractive-index difference ($\Delta n$) is insignificant. The negative influence of $\Delta n$ can be minimized by making the predominant orientation of the grains coincide with the direction of the optic axis. It is well known that the grain boundaries have a significant effect on the other properties of ceramics,[7] and therefore the characteristics of ceramics and single crystals usually differ.

Zinc oxide manifests typical semiconductor properties, possessing in this case a large fraction of ionic bonding. The characteristics of zinc oxide are most fully presented in a recent review.[8] Under ordinary circumstances, ZnO possesses the hexagonal wurtzite structure, in which each $O^{2-}$ ion is surrounded by a tetrahedron composed of four $Zn^{2+}$ ions. The crystal-structure constants have the following values: $a=3.2497$ Å and $c=5.2069$ Å, and their ratio of $c/a = 1.602$ is close to ideal. The refractive-index difference of zinc oxide oscillates within the limits 0.016–0.018 in the visible region and equals 0.009 at 405 nm. There are numerous forms of ZnO: single crystals, thin films and filaments, nanocrystals, needles, etc., and, as a rule, two emission bands are recorded: a short-wavelength one close to the absorption edge of the crystal—i.e., edge luminescence—and a long-wavelength (green) band, which we call intraband luminescence. Edge luminescence has an exciton nature, while the intraband luminescence is associated with the presence of oxygen or zinc vacancies or with residual impurities.[8,9]

The prospects of using ZnO and ZnO:Ga in short-wavelength optoelectronics and laser and scintillation engineering have been considered.[1–5] It is important in the latter case that ZnO possesses a relatively small (for scintillators) band gap ($E_g=3.37$ eV), since the conversion efficiency of scintillators increases as the band gap decreases. Moreover, the exciton binding energy (60 meV) in ZnO is greater by a factor of 2.4 than the thermal energy $kT$ (for $T \approx 290$ K). As a result, the edge luminescence has high intensity at room temperature. Edge luminescence is the most important in the case of rapid scintillators, since it is characterized by nanosecond and subnanosecond de-excitation times. The following characteristics of ZnO are also important for scintillation detectors: transparency in the visible region, good thermal and mechanical properties, fairly high density (5.61 $g/cm^3$) and high radiation stability.[10]

This paper presents the first results on the synthesis and study of ZnO and ZnO: Ga scintillation ceramics. Uniaxial hot pressing was used to obtain the ceramics. This method was used earlier when developing the scintillation ceramics $Gd_2O_2S$: Pr, Ce (Ref. 11) and ZnO: Zn.[12] The optical, luminescence, scintillation, and other properties of the resulting ZnO and ZnO: Ga ceramics are studied here.

### EXPERIMENTAL TECHNIQUE

All the ceramics were obtained in the form of disks 24 mm in diameter and 1.5 mm thick (after polishing). Both the pure and the doped ceramics were synthesized from two forms of the starting zinc oxide: commercial ZnO powders of VHP grade (designated ZnO I) produced by ZAO NPF Lyuminofor (Stavropol) and a specially purified product of the firm Alfa Aesar (designated ZnO II). Powdered $Ga_2O_3$ and $Ga(NO_3)_3$, also produced by Alfa Aesar, were used for doping. Based on an earlier test of the synthesis of ceramics and data on powdered ZnO: Ga,[5,8] the gallium concentration in ZnO was chosen within the limits from 0.05 to 0.1 wt% (in what follows, the impurity concentration is indicated everywhere in weight percent). Gallium was introduced in the form of $Ga_2O_3$ and $Ga(NO_3)_3$, corresponding to the samples designated as ZnO: Ga or ZnO: Ga,N. Unlike gallium, which forms donor levels in ZnO, nitrogen forms shallow acceptor levels.[5] It is assumed that, because of donor-acceptor recombination, the edge band in the samples doped with gallium and nitrogen must be shifted toward longer wavelengths by comparison with those in ZnO: Ga.

The morphology and mean size of a powder grain and of the ceramic samples was studied by means of optical microscopy. The lattice parameters and the degree of texturing of the ceramic were studied on a DRON-2 x-ray diffractometer with a copper anode and a nickel filter when the x-ray reflections were recorded on a recorder chart. To determine the parameters of the crystal cell, a system of crystallographic planes (105) and (300) with large reflection angles $2\theta$ (104.12° and 110.52°) was chosen. A technique proposed in Ref. 13 for calculating the texture factor was used to monitor the texture—i.e., the predominant orientation of the crystallographic planes in the ceramic samples. The texture factor was calculated from unambiguously indexed reflection lines from the (100), (002), (101), (102), (110), and (103) planes by comparing the intensities of these lines from the texturized test sample with standard data for this material, taken from the I.C.P.D.S. card file. The total spectral transmittance of the samples was determined on a Hitachi-330 spectrophotometer equipped with an attachment having an integrating sphere 60 mm in diameter.

The x-ray-luminescence spectra were measured using an x-ray tube with a copper anode, operating in the 55-kV, 40-mA regime. An Al filter 3 mm thick was used to cut off the soft component of the x-ray emission. The emission spectra were measured by means of an Acton Research Co. VM-504 monochromator (diffraction grating 1200 line/mm). A Hamamatsu R934-04 photomultiplier was used as a photodetector. All the measured spectral curves were corrected taking into account the photomultiplier sensitivity and the monochromator transmissivity for various wavelengths.

The absolute luminous yield of single-crystal $BaF_2$ was measured using a Hamamatsu R1791 photomultiplier by comparing the position of the maximum of the photopeak for the $Cs^{137}$ spectrum (662 keV) with the position of the center of gravity of the single-electron spectrum.[14] The luminous yield of standard single-crystal $BaF_2$ was equal to 8880 photons per MeV. For the ceramic samples of ZnO and ZnO: Ga, it was not possible to reliably resolve the position of the photopeak using excitation with $Cs^{137}$ gamma quanta. The absolute luminous yield of the resulting ceramic was determined by the ratio of its overall intensity to the x-ray-luminescence intensity of $BaF_2$, multiplied by the absolute luminous yield of barium fluoride.

$$L = \left( \int I_{ZnO}(\lambda)d\lambda \bigg/ \int I_{BaF_2}(\lambda)d\lambda \right) 8880 \text{ photon/MeV}$$

The luminescence kinetics were measured using an original experimental apparatus.[15] An x-ray source with pulse width $\leq 1$ ns, operating at a voltage of 30 kV and a maximum current amplitude of 500 mA, was used to excite the samples. The recording layout consisted of an FÉU 71 photomultiplier, operating in the photon-counting regime, and the accompanying electronics, which allowed measurements to be made in various time ranges with a resolution no worse than 100 ps.

### OPTICAL AND X-RAY CHARACTERISTICS OF THE SAMPLES

The original ZnO powders were characterized by a fairly homogeneous grain composition both in morphology and in size. The grains were predominantly round and close to isometric in shape, with a size of about 1–3 $\mu$m. The results of the x-ray structural analysis showed that the lattice parameters of the original powders correspond to those for single crystals: $a = 3.2497$ Å and $c = 5.2069$ Å. The hot pressing process has a significant effect on the lattice parameters: A decrease of the $c$ parameter to 5.2034–5.2043 Å was observed for the undoped ZnO ceramics. The introduction of a dopant impurity with smaller ionic radius ($r[Ga^{3+}] = 0.47$ Å, and $r[Zn^{3+}] = 0.6$ Å) into the ZnO lattice did not appreciably change the parameters of the ZnO: Ga ceramic. When gallium nitrate was introduced into the ZnO, parameter $c$ decreased. A large decrease of parameter $a$ was also noted, which is observed in ceramic samples when there is a minimum dopant concentration, regardless of the type of compound in the form of which it was introduced.

Studies of the predominant orientation of the principal crystallographic planes showed that all the ceramics are characterized by texture along the (100) and (110) planes of the prism and the (101) planes of the rhombohedron. The manifestation of texture over the three planes is a consequence of the fact that it spread (the angle between the planes does not exceed 30°). The predominant orientation is observed along the (101) plane of the rhombohedron for the undoped ZnO ceramic. The texture over the planes of the prism predominates in the gallium-doped samples. As a whole, the value of the texture does not exceed 0.3.

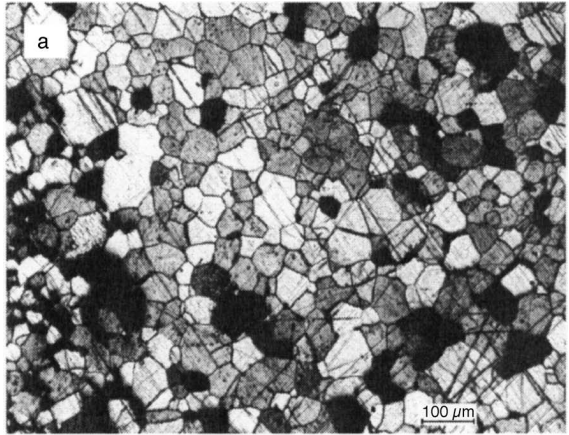

FIG. 1. Microstructure of ZnO II (a) and ZnO:Ga II ceramics (b).

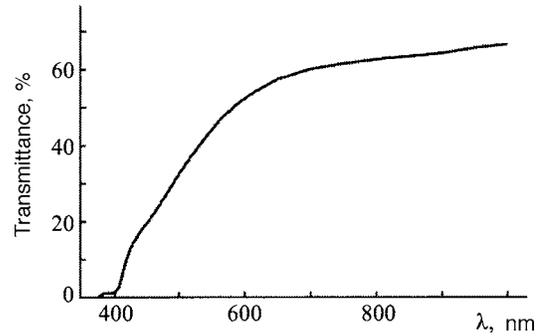

FIG. 2. Total transmission spectrum of ZnO II ceramic 1.5 mm thick.

Figure 1 shows photographs of typical microstructures of the ZnO II and ZnO: Ga II ceramics. The character of the represented microstructures reflects, first, the intense progress of recrystallization processes during the formation of the ZnO ceramic, as a result of which the grain size increases by 1.5-2 orders of magnitude by comparison with the size of the original particles, and, second, the inhibiting role of the dopant, a consequence of which is that the recrystallization processes ceases. An analysis of the data in Fig. 1 showed that the grain size is 30–90 $\mu$m in the undoped sample and 15–35 $\mu$m in ZnO: Ga II.

The density of all the ceramic samples was more than 0.99 relative to the x-ray-structural density of ZnO. The resulting samples had a characteristic color. The ZnO I samples were an intense red, but the ZnO II had a yellow-red tint. The doped ceramic was in between light blue and dark blue, whose intensity increased with increasing gallium concentration. The highest transparency characterizes the ZnO II ceramic, whose total transmission spectrum is shown in Fig. 2. Absorption in the short-wavelength region of transparency reduces the transmission level of the ZnO II ceramic in this region because of the presence of color (the cause of which will be explained). A similar phenomenon for single-crystal ZnO (the presence of red coloration and the shift of the short-wavelength transmission limit toward longer wavelengths) was noted in Ref. 16 and was explained by the presence of a high concentration of oxide vacancies. High-temperature annealing in an oxygen atmosphere allowed the authors to decolor the ZnO single crystals, and this was accompanied by an increase of transparency in the entire visible region and a corresponding shift of the transmission limit toward short wavelengths.

It should be pointed out that the resulting transparency level of the undoped ceramic in the long-wavelength region ($\lambda \geq 600$ nm) approaches that for single-crystal ZnO.[17] It is essential in this case that such a result is achieved for a ceramic in which the predominant orientation of the grains is not optimal for a hexagonal structure, since it does not coincide with the direction of the optic axis.

## THE LUMINESCENCE AND SCINTILLATION CHARACTERISTICS OF THE CERAMICS

Figure 3a shows the x-ray-luminescence spectra of ZnO I and ZnO II ceramics in comparison with the spectrum of the traditional scintillator $BaF_2$. It can be seen that intraband luminescence, i.e., a wide band with a maximum at 520 nm, predominates in the undoped ceramics. The ZnO I ceramic manifests a weak edge luminescence (Fig. 3b), whereas the edge luminescence in the original powders and single crystals is fairly intense.[18,19] The luminous yield of ZnO II was greater than that of $BaF_2$ and substantially greater than that of ZnO I.

The luminescence spectra of the doped ceramics are shown in Fig. 4. The intensities of both luminescence bands increase in the transition I→II (Figs. 4a and 4b), and this is valid both for both the ZnO: Ga and the ZnO: Ga,N samples. When the gallium concentration is increased from 0.1% (Fig. 4c), the luminous yield decreases by comparison with that for ZnO: Ga,N II (0.075%) (Fig. 4b). The radiation in the ZnO: Ga ceramics is concentrated predominantly in the short-wavelength region (Fig. 4d), whereas the intraband luminescence with a maximum at 510 nm predominates in ZnO: Ga,N (Figs. 4a–4c). It should be pointed out that the luminous yield of the ZnO: Ga II and ZnO: Ga,N II ceramics is identical, but the presence of nitrogen in the sample strengthens the intraband luminescence and weakens the edge luminescence. The position of the maximum of the edge-luminescence band was identical in ZnO: Ga and ZnO:

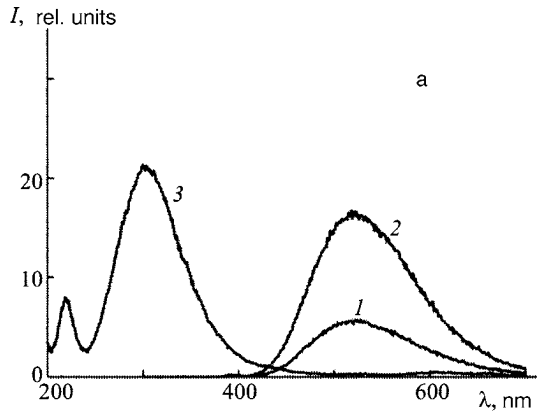

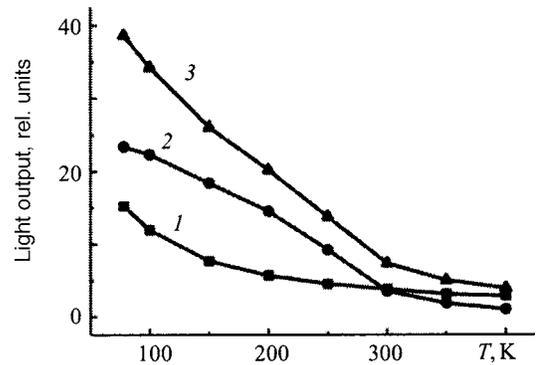

FIG. 5. Temperature dependences of the intensities of *1*—edge luminescence (497.5 nm), *2*—intraband luminescence (510 nm), and *3*—total luminescence of ZnO: Ga II ceramic.

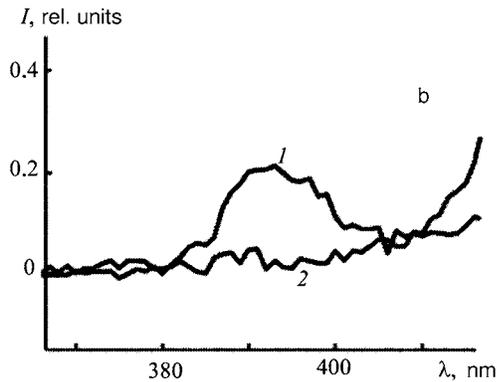

FIG. 3. (a) Luminescence spectra of undoped ceramics ZnO I—*1*, ZnO II—*2* and single-crystal BaF$_2$—*3* at room temperature. (b) Segment of spectra at an enlarged intensity scale.

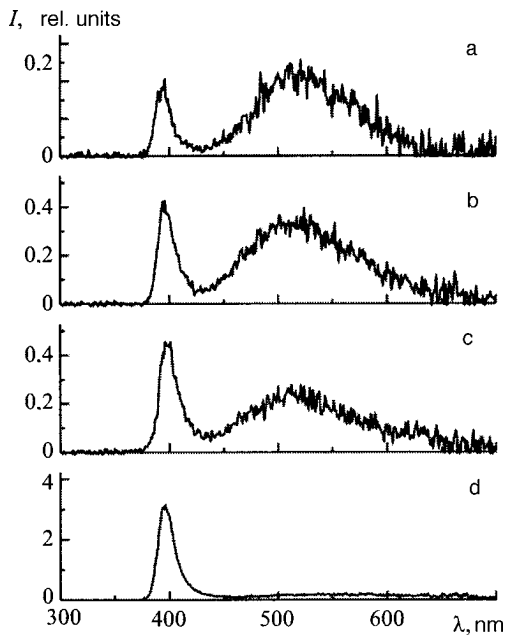

FIG. 4. Luminescence spectra of ZnO: Ga ceramics at room temperature. (a) ZnO: Ga,N I (0.075%), (b) ZnO: Ga,N II (0.075%), (c) ZnO: Ga,N II (0.1%), (d) ZnO: Ga II (0.075%).

Ga,N: 397.5 nm at 300 K; i.e., the expected long-wavelength shift when an acceptor (N) was introduced was not observed. Moreover, the maximum of the edge band in the doped samples (Fig. 4) is shifted relative to the maximum in ZnO I (Fig. 3b) by only about 10 meV.

The intraband luminescence of ZnO can be associated with zinc vacancies $V_{Zn}$,[8] oxygen vacancies $V_O$,[8,9] antinode zinc $O_{Zn}$,[20] and other centers. A recent special study[18] showed that $V_{Zn}$ centers are responsible for the luminescence band with the maximum at 2.35 eV that is recorded in our ceramics, while oxygen vacancies result in shorter-wavelength radiation (2.53 eV). Obviously, the introduction of $Ga_2O_3$ and ZnO in our case reduces the number of zinc vacancies in the sample and strengthens the edge luminescence.

Figure 5 shows how the luminescence intensity depends on temperature for a sample of ZnO: Ga II. The intensity of the edge luminescence at 78 K is a factor of 3 greater than at 300 K. For $T > 300$ K, the intensity of the edge band becomes greater than that of the intraband luminescence.

Examples of the kinetic curves for samples of ZnO: Ga and ZnO: Ga,N are shown in Fig. 6. A rapid falloff of the x-ray luminescence predominates n the samples. The rapid component in ZnO: Ga with a falloff constant of about 1 ns occupies two decimal places in intensity. The ZnO: Ga,N ceramic possesses a hyperbolic luminescence falloff, and a section with a falloff constant of $3.3 \pm 0.3$ ns can be distinguished at the beginning of the curve (Fig. 6b). The ZnO: Ga ceramic is thus preferable to ZnO: Ga,N in response rate.

The main characteristics of the ceramics are shown in Table I. The luminous yield of the pure sample of ZnO II is fairly high: 9050 photon/MeV. The doped ceramics have a low luminous yield: 420 photon/MeV. However, this is the so-called technical output, associated with the low transparency of the ceramics so far obtained. The physical luminous yield must be greater, since it is very high in the initial ZnO: Ga. Moreover, it should again be emphasized that these data relate to samples that have not undergone heat treatment. These data need to be regarded as preliminary, since heat treatment in a special atmosphere is an integral procedure for forming scintillators not only on the basis of ZnO.[5,11] Our exploratory experiments on the heat treatment of ceramic samples of ZnO: Ga showed that the intensity of the edge luminescence becomes somewhat greater in an annealed ceramic than in the original sample, with the transparency of the ceramic also increasing slightly.

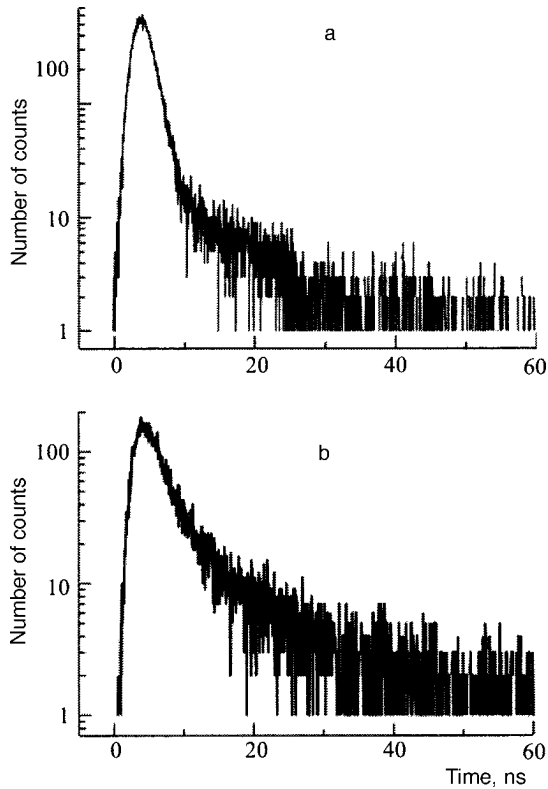

FIG. 6. Luminescence kinetics of ZnO: Ga II (a) and ZnO: Ga,N II (b) ceramics.

From the viewpoint of luminous yield, the optimum gallium concentration (when it is introduced in the form of $Ga_2O_3$) in ZnO was 0.075%, and the result is found to be in agreement with the data for powdered[5] and thin-film[21] ZnO: Ga. An increase of the gallium concentration by 30% resulted in a decrease of the luminous yield by almost a factor of 2. Concentration quenching of the edge luminescence in ZnO: Ga thus begins at a gallium concentration that exceeds 0.075%. Similar values are characteristic of the ceramic scintillators $Gd_2O_2S$: Pr,Ce.[11] This is a fairly low value, since the optimum dopant concentration is 0.1–0.2% in traditional scintillators (NaI: Tl, CsI: Tl). It should be emphasized that a decrease of the optimum dopant concentration in a ceramic by comparison with the powdered analogs was observed by the authors using such luminescent ceramics as ZnS:Cu as examples. This feature of the ceramics is probably caused by the higher solubility of dopant impurities in the matrix lattice during hot pressing.

## CONCLUSION

Uniaxial hot pressing has been used to obtain ceramics based on zinc oxide, and their optical, x-ray-structural, luminescence, and scintillation characteristics have been studied. It has been shown that, by changing the concentration of the dopant (Ga) and the codopant (N), it is possible to change the intensity of the edge band and the intraband luminescence of ZnO, as well as their ratio. The resulting ceramics possess short luminescence-falloff times.

Undoped ZnO II ceramic has good transparency in the visible region and a fairly high luminous yield, unlike ZnO: Ga ceramic. Based on the first data obtained by the authors, they have developed a strategy for improving the quality of the ZnO: Ga ceramics (producing, most importantly, increased transparency of the samples), which should improve the scintillation characteristics of ceramics based on zinc oxide.

a)Email: E.Gorokhova@rambler.ru

TABLE I. Comparison of the characteristics of ceramics based on zinc oxide and a $BaF_2$ standard scintillator.

| Sample | Light, yield, photon, MeV | Luminescence maximum basic/suppl., nm | Falloff time, ns |
| --- | --- | --- | --- |
| $BaF_2$ | 8880[a] | 310/220 | 0.88/600 |
| ZnOI | 3200 | 520 | ≈ 10 |
| ZnOII | 9050 | 520 | ≈ 10 |
| ZnO:GaII (0.075%) | 420 | 395/510 | ≈ 1.0 |
| ZnO:Ga, NII (0.075%) | 420 | 395/510 | 3.3 ± 0.3 |

[a] The luminous yield is obtained for the available $BaF_2$ crystal, having the same size as the test ceramic. The best samples of $BaF_2$ scintillators possess a luminous yield of up to 11 000 photons per MeV.

[1]S. E. Derenzo, M. J. Weber, and M. K. Klintenberg, "Temperature dependence of the fast, near-band-edge scintillation from CuI, $HgI_2$, $PbI_2$, ZnO: Ga and CdS: In," Nucl. Instrum. Methods Phys. Res. A **486**, 214 (2002).
[2]P. J. Simpson, R. Tjossem, A. W. Hunt, K. G. Lynn, and V. Munne, "Superfast timing performance from ZnO scintillators," Nucl. Instrum. Methods Phys. Res. A **505**, 82 (2003).
[3]M. Katagiri, K. Sakasai, M. Matsubayashi, T. Nakamura, Y. Kondo, Y. Chujo, H. Nanto, and T. Kojima, "Scintillation materials for neutron-imaging detectors," Nucl. Instrum. Methods Phys. Res. A **529**, 274 (2004).
[4]N. Kubota, M. Katagiri, K. Kamijo, and H. Nanto, "Evolution of ZnS-family phosphors for neutron detectors using counting method," Nucl. Instrum. Methods Phys. Res. A **529**, 321 (2004).
[5]E. D. Bourret-Courchesne, S. E. Derenzo, and M. J. Weber, "Semiconductor scintillators ZnO and $PbI_2$: co-doping studies," Nucl. Instrum. Methods Phys. Res. A **579**, 1 (2007).
[6]S. J. Duclos, C. D. Greskovich, R. J. Lyons, J. S. Vartuli, D. M. Hoffman, R. J. Riener, and M. J. Lynch, "Development of the HiLight scintillator for computed tomography medical imaging," Nucl. Instrum. Methods Phys. Res. A **505**, 68 (2003).
[7]Y. Sato, T. Yamamoto, and Y. Ikuhara, "Atomic structures and electrical properties of ZnO grain boundaries," J. Am. Ceram. Soc. **90**, 337 (2007).
[8]U. Orgur, Ya. I. Alivov, C. Liu, A. Teke, M. A. Reshnikov, S. Dogan, V. Avrutin, S.-J. Cho, and H. Morkoc, "A comprehensive review of ZnO materials and devices," J. Appl. Phys. **98**, 041301 (2005).

[9] Ya. I. Alivov, M. V. Chukichev, and V. A. Nikitenko, "Green luminescence band of zinc oxide films doped with copper during thermal diffusion," Fiz. Tekh. Poluprovodn. **38**, 34 (2004) [Semiconductors **38**, 31 (2004)].

[10] S. O. Kucheyev, J. S. Williams, C. Jagadish, J. Zou, C. Evans, A. J. Nelson, and A. V. Hamza, "Ion-beam-produced structural defects in ZnO," Phys. Rev. B **67**, 094115 (2003).

[11] E. I. Gorokhova, V. A. Demidenko, S. B. Eron'ko, S. B. Mikhrin, P. A. Rodnyĭ, and O. A. Khristich, "Spectrokinetic characteristics of $Gd_2O_2S$:Pr,Ce ceramics," J. Opt. Technol. **73**, No. 2, 71 (2006) [J. Opt. Technol. **73**, 130 (2006)].

[12] V. A. Demidenko, E. I. Gorokhova, I. V. Khodyuk, O. A. Khristich, S. B. Mikhrin, and P. A. Rodnyi, "Scintillation properties of ceramics based on zinc oxide," Radiat. Meas. **42**, 549 (2007).

[13] Kunio Matsuzaki, Akishisa Jnoue, and T. Masumoto, "Oriented structure and semiconductor properties in dense $Y_1B_2Cu_3$ oxides prepared by Press Forging," Jpn. J. Appl. Phys. **29**, 1789 (1990).

[14] J. T. M. De Haas, P. Dorenbos, and C. W. E. van Eijk, "Measuring the absolute light yield of scintillators," Nucl. Instrum. Methods Phys. Res. A **537**, 97 (2005).

[15] P. A. Rodnyi, S. B. Mikhrin, A. N. Mishin, and A. V. Sidorenko, "Small-size pulsed x-ray source for measurements of scintillator decay time constants," IEEE Trans. Nucl. Sci. **48**, 2340 (2001).

[16] A. Mycielski, L. Kowalczyk, A. Szadkowski, B. Chwalisz, A. Wysmolek, R. Stepniewski, J. M. Baranowski, M. Potemski, A. Witowski, R. Jakiela, A. Barcz, B. Witkowska, W. Kaliszek, A. Jedrzejczak, A. Suchocki, E. Lusakowska, and E. Kaminska, "The chemical-vapor-transport growth of ZnO single crystals," J. Alloys Compd. **371**, 150 (2004).

[17] L. Xin-Hua, X. Jia-Yue, J. Min, S. Hui, and L. Xiao-Min, "Electrical and Optical properties of bulk ZnO single crystal grown by flux Bridgman method," Chin. Phys. Lett. **23**, 3356 (2006).

[18] T. Moe Berseth, B. G. Svenson, A. Yu. Kuznetsov, P. Klason, Q. X. Zhao, and M. Willander, "Identification of oxygen and zinc vacancy optical signal in ZnO," Appl. Phys. Lett. **89**, 262112 (2006).

[19] P. A. Rodnyĭ, G. B. Stryganyuk, and I. V. Khodyuk, "Luminescence of a ZnO:Ga crystal excited in the vacuum UV," Opt. Spectrosc. **104**, 257 (2008) [Opt. Spectrosc. **104**, 210 (2008)].

[20] B. Lin, Z. Fu, and Y. Jia, "Green luminescent center in undoped zinc oxide films deposited on silicon substrates," Appl. Phys. Lett. **79**, 943 (2001).

[21] T. Makino, Y. Segawa, S. Yoshida, A. Tsukazaki, A. Ohtomo, and M. Kawasaki, "Gallium concentration dependence of room-temperature near-band-edge luminescence in $n$-type ZnO:Ga," Appl. Phys. Lett. **85**, 759 (2004).